\documentclass[twocolumn,prl,amsmath,amssymb,superscriptaddress,showpacs]{revtex4}

\usepackage{graphicx}
\usepackage{dcolumn}
\usepackage{bm}

\def\be{\begin{eqnarray}}
\def\ee{\end{eqnarray}}
\def\reff#1{(\ref{#1})}

\newcommand{\ten}[1]{{\stackrel{=}{#1}}}

\newcommand {\vH}{{\bf H}}

\newcommand {\vM}{{\bf M}}

\newcommand {\vI}{{\bf I}}

\newcommand {\vz}{\hat{\bf z}}
\newcommand {\vx}{\hat{\bf x}}
\newcommand {\vy}{\hat{\bf y}}
\newcommand {\ver}{{\bf r}}

\newcommand {\vn}{{\bf n}}
\newcommand {\vm}{{\bf m}}
\newcommand {\vk}{{\bf k}}

\begin{document}


\title{Unified Homogenization Theory for Magnetoinductive and Electromagnetic Waves in Split Ring Metamaterials}

\author{J. D. Baena}
\email{juan_dbd@us.es} 
\affiliation{Departamento de Electr\'onica y Electromagnetismo, Universidad de Sevilla, 41012-Sevilla, Spain}
\affiliation{Departamento de F\'isica Aplicada, Universidad Nacional de Colombia, Bogot\'a, Colombia}

\author{L. Jelinek}
\email{l_jelinek@us.es} 
\affiliation{Departamento de Electr\'onica y Electromagnetismo, Universidad de Sevilla, 41012-Sevilla, Spain}

\author{R. Marqu\'es}
\email{marques@us.es} 
\affiliation{Departamento de Electr\'onica y Electromagnetismo, Universidad de Sevilla, 41012-Sevilla, Spain}

\author{M. Silveirinha}
\email{mario.silveirinha@co.it.pt}
\affiliation{Departamento de Engenharia Electrot\'{e}cnica, Universidade de Coimbra, 3030-Coimbra, Portugal}


\begin{abstract}
A unified homogenization procedure for split ring metamaterials taking into account time and spatial dispersion is introduced. The procedure is based on two coupled systems of equations. The first one comes from an approximation of the metamaterial as a cubic arrangement of coupled LC circuits, giving the relation between currents and local magnetic field. The second equation comes from macroscopic Maxwell equations, and gives the relation between the macroscopic magnetic field and the average magnetization of the metamaterial. It is shown that electromagnetic and magnetoinductive waves propagating in the metamaterial are obtained from this analysis. Therefore, the proposed time and spatially dispersive permeability accounts for the characterization of the complete spectrum of waves of the metamaterial. Finally, it is shown that the proposed theory is in good quantitative and qualitative agreement with full wave simulations.
\end{abstract}


\maketitle

Diamagnetic properties of systems of conducting rings are known since long by physicists. In 1852 Wilhem Weber \cite{Weber-1852} tried to explain natural diamagnetism (discovered by Faraday some years before) as a consequence of the excitation of induced currents in some hypothetical conducting loops that supposedly existed in diamagnetic materials. In order to enhance the magnetic properties of artificial media (or metamaterials in modern terminology) made from metallic conducting rings, S. A. Shelkunoff proposed in 1952 to introduce a capacitor \cite{Shelkunoff-1966}, so as the rings become resonant. More recently J. B. Pendry et al. \cite{Pendry-1999} proposed to replace the capacitively loaded rings by planar split ring resonators (SRRs) which substitute the lumped capacitor by a distributed capacitance between the rings. Because Pendry's SRRs can be easily manufactured by using standard printed circuit technologies, this design opened the way to manufacturing true magnetic metamaterials made of many individual elements (SRRs) at many laboratories around the world. As a consequence of its resonant behavior, capacitively loaded rings and/or SRRs can produce metamaterials with negative magnetic permeability above resonance. It is also well known \cite{Smith-2000, Shelby-2001, Marques-2002} that when a system of these elements is properly combined with another system of elements (metallic wires or plates, for instance) producing a negative electric permittivity \cite{Rotman-1968}, a metamaterial with simultaneously negative permittivity and permeability (or left-handed metamaterial \cite{Veselago-1968}) arises in the frequency band were both subsystems present negative parameters. Remarkably, the electric and magnetic properties of such combinations are, quite approximately, the superposition of the electric and magnetic properties of each subsystem. This superposition hypothesis is not apparent at all, since the elements of both subsystems must be placed closely in space, and therefore electromagnetic coupling may be present. In fact, this has been a controversial issue (see for instance \cite{Pokrovsky-2002, Marques/Smith-2004}). For the specific SRRs and wires configuration proposed in \cite{Smith-2000} this hypothesis has been recently demonstrated by one of the authors \cite{Silveirinha-2007} and, in general, it can be admitted that the aforementioned superposition hypothesis is valid provided the elements of both subsystems are placed in such a way that their quasi-static fields do not interact or interact weakly \cite{Marques/Smith-2004}. Almost simultaneously, other analyses and experiments \cite{Shamonina-2002, Belov-2005} did show that SRR based metamaterials also support, in some frequency bands, slow waves based on short range interactions between the SRRs; the so called magnetoinductive (MI) waves, which can not be deduced from the usually assumed local magnetic permeability of the metamaterial. Interestingly, many of the physical effects expected in negative permeability and left-handed metamaterials, such as frequency band gaps and frequency bands of backward-wave propagation, also come out from the analysis when the coupling between electromagnetic and MI waves in SRR systems is considered \cite{Syms-2005}, thus providing an alternative explanation for such effects. Although the analysis in \cite{Syms-2005} has a great heuristic value, it can not be considered as fully satisfactory because it only considers one-dimensional systems in the nearest neighbors approximation. On the other hand, the presence of waves which can not be deduced from a local time dispersive magnetic permeability in split ring metamaterials can be expected from the fact that its periodicity is usually not smaller than one tenth of a wavelength. As it is well known \cite{Landau-8}, when the periodicity of a given medium approaches the wavelength of the electromagnetic radiation, it becomes not only time dispersive but also spatially dispersive. Therefore, it can be expected that both, electromagnetic and MI waves, would come out from the analysis if spatial dispersion in split ring metamaterials were taken into account. In fact, the main purpose of this paper is to show that both kinds of waves can be obtained from a unified analysis when a proper homogenization procedure, taking into account spatial dispersion, is developed.

\begin{figure}
\centering
\includegraphics[width=0.8\columnwidth]{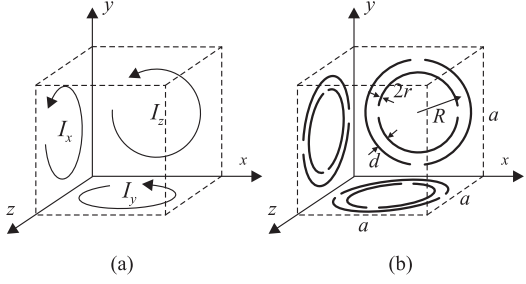}
\caption{\label{Fig1} (a): Unit cell of a material formed by a cubic array of current loops. Each unit cell has three current loops centered at the faces of the cube. (b): Similar to (a) but for a realistic metamaterial formed by edge coupled SRRs with two splits. The SRRs are formed by circular wires with radius $r$. The distance between the inner and outer rings is $d$ and the average radius of the particle is $R$.}
\end{figure}

In order to simplify the analysis, we will consider an \emph{ideal} metamaterial made of a cubic arrangement of LC circuits supporting current loops as sketched in Fig.1a for a unit cell. We define the \emph{current vector} $\vI^{\vn}$ on each unit cell as $\vI^{\vn}=(I_x^{\vn}, I_y^{\vn}, I_z^{\vn})$, where $\vn=(n_x, n_y, n_z)$ specifies the location of each unit cell in the lattice and $I_i^{\vn}$ denotes the current along the loop located in the face normal to the $i$-direction of unit cell of index $\vn$. The time dependence is assumed of the form $\vI^{\vn}\propto\exp(j\omega t)$. Each component of the current vector is governed by equation
\begin{equation}
  \label{I-equation}
  \left(j\omega L + \frac{1}{j\omega C}\right) I_i^{\vn} = -j\omega\Phi_i^{\vn},
\end{equation}
where $L$, $C$ are self-inductance and self-capacitance of the circuit (losses are neglected by simplicity, although they can be easily introduced in the analysit through a ring resistance). In Eq. \reff{I-equation} $\Phi_i^{\vn}$ is the total magnetic flux through the considered loop which, using Lorentz local field approximation, can be calculated as 
\begin{equation}
  \label{Phi}
  \Phi_i^{\vn} = A\mu_0 \left( H_i+\frac{M_i}{3} \right) + \sum_{({\vm}\ne {\vn}) \wedge (j \ne i)}^{r<R} M_{ij}^{\vn \vm} I_j^{\vm},
\end{equation}
where $A$ is the area of the loop, $\vH$, $\vM$ are the macroscopic magnetic field and magnetization on the ring, and $M_{ij}^{\vn \vm}$ are the mutual inductances between the loops with indexes $\vn$ and $\vm$. The summation is extended only to a sphere around the unit cell $\vn$ which contains several unit cells but which radius $R$ is smaller than the wavelength.

In the following we will assume a spatial field dependence of the kind $\{\vH, \vM\} = \{\vH_0, \vM_0\}\exp(-j\vk\cdot\ver)$ and $\vI^{\vn} = \vI_0\exp(-ja\vk\cdot\vn)$, where $a$ is the lattice periodicity and $\vI_0$, $\vH_0$, $\vM_0$ are constant vectors. With the assumed time and space dependence, the macroscopic Maxwell equations lead to 
\begin{equation}
  \label{H-M-equation}
  \left(k_m^2 - k^2 \right)\vH_0 + \left(k_m^2 - \vk\vk\cdot \right) \vM_0 = 0,
\end{equation}
with $k_m=k_0\sqrt{\varepsilon_r}=\omega\sqrt{\varepsilon_r\varepsilon_0\mu_0}$, where $\varepsilon_r$ is the \emph{macroscopic} relative dielectric constant of the metamaterial. By combining \reff{I-equation} and \reff{Phi} the following equation for $\vI_0$ is obtained 
\begin{equation}
  \label{I0-equation}
  \ten{Z}(\vk, \omega) \cdot\vI_0 = -j\omega A\mu_0 \left(\vH_0 + \frac{\vM_0}{3} \right),
\end{equation}
where $\ten{Z}(\vk, \omega)$ is an impedance matrix which incorporates all the magneto-inductive effects between the neighboring rings. Explicit expressions for the diagonal and the off-diagonal terms of $\ten{Z}$ are 
\begin{equation}
  \label{Z-diagonal}
  Z_{ii} = j\omega L\left\{ 1 - \frac{\omega_0^2}{\omega^2} + \sum_{\vn\ne \bf{0}}^{r<R} \frac{M_{ii}^{\bf{0}\vn}}{L} e^{-ja\vk\cdot\vn}\right\}
\end{equation}
\begin{equation}
  \label{Z-cross}
  Z_{ij} = Z_{ji} = j\omega L \sum_{\vn}^{r<R} \frac{M_{ij}^{\bf{0}\vn}}{L} e^{-ja\vk\cdot\vn}\;;\;\; i\ne j
\end{equation}
where $\omega_0=1/\sqrt{LC}$ is the frequency of resonance of the rings, with similar expressions for the remaining components of $\ten{Z}$. Taking into account that $\vM_0 = A\vI_0/a^3$, it is possible to combine \reff{H-M-equation} with \reff{I0-equation} which gives 
\begin{equation}
  \label{d-r}
  \left\{ \ten{Z}(\vk,\omega) - \frac{j\omega\mu_oA^2}{3a^3} \frac{2k_m^2+k^2-3\vk\vk}{k_m^2-k^2} \right\} \cdot \vI_0 = 0.
\end{equation}
The dispersion equation for plane waves in the metamaterial can be obtained by equating the determinant of \reff{d-r} to zero. In the most general case this equation can be only solved numerically. However, in some cases, it is possible to give analytical solutions. In particular, for propagation along one of the coordinate axis (for $\vk=k\vx$, for instance) the summation in \reff{Z-cross} vanishes because all unit cells in planes perpendicular to the $x-$axis are in phase and mutual inductances cancel out couple by couple (for instance, the mutual inductances between the ring marked $I_x$, and the rings placed on the top and lower faces of the cube of Fig.1.(a) cancel each other, and so on). Therefore, the matrix $\ten{Z}(\vk, \omega)$ becomes diagonal, and \reff{d-r} can be easily solved. This gives two branches: a longitudinal wave with $\vI_0 = I_{0x}\vx$ given by 
\begin{equation}
  \label{l-wave}
  Z_{xx} - \frac{2}{3} \frac{j\omega\mu_oA^2}{a^3} =0
\end{equation}
and a transverse wave with $\vI_0 = I_{0y}\vy + I_{0z}\vz$ given by 
\begin{equation}
  \label{t-wave}
  Z_{yy} - \frac{j\omega\mu_0A^2}{3a^3}\frac{2k_m^2 + k_x^2}{k_m^2-k_x^2} = 0.
\end{equation}
If only interactions with the closest rings are considered for the computation of the summations in \reff{Z-diagonal} and \reff{Z-cross} the dispersion equation for the longitudinal wave becomes 
\begin{equation}
  \label{l-MI}
  \frac{\omega_0^2}{\omega^2} = 1+ 2\frac{M_a}{L}\cos(ak_x) + 4\frac{M_c}{L}-\frac{2}{3}\frac{\alpha_0}{a ^3},
\end{equation}
where $\alpha_0=\mu_0A^2/L$ and $M_a$ and $M_c$ are the mutual inductances between closest rings of the same orientation, placed in the coplanar and the axial directions, respectively (as it was already mentioned, the contributions of the mutual inductances between rings of different orientations cancel each other for this particular propagation). It can be easily recognized in \reff{l-MI} the dispersion relation for longitudinal magnetoinductive waves \cite{Shamonina-2002}, with some small corrections, which take into account the effects of the rings other than the nearest neighbors in the axial direction. In the same approximation the dispersion equation for transverse waves can be written as 
\begin{equation}
  \label{t-wave-2}
  \frac{k_x^2}{k_m^2}-1 = \frac{\frac{\alpha_0}{a^3}}{\frac{\omega_0^2}{\omega^2} - 1- \frac{2M_c}{L}\cos(ak_x)-\frac{2(M_a+M_c)}{L}-\frac{\alpha_0}{3a^3}}.
\end{equation}
For high values of $k_x$ ($k_x\gg k_m$) this equation reduces to
\begin{equation}
\label{t-MI}
\frac{\omega_0^2}{\omega^2} = 1 + \frac{2M_c}{L}\cos(ak_x) + \frac{2(M_a+M_c)}{L} + \frac{\alpha_0}{3a^3},
\end{equation}
which corresponds to the dispersion relation for transverse magnetoinductive waves \cite{Shamonina-2002}. On the other hand, in the long wavelength limit ($ak_x\ll 1$) \reff{t-wave-2} reduces to 
\begin{equation}
  \label{t-EM}
  \chi(\omega) = \frac{k_x^2}{k_m^2} -1 = \frac{\frac{\alpha_0}{a^3}}{\frac{\omega_0^2}{\omega^2} - 1 - \frac{2M_a}{L}- \frac{4M_c}{L}- \frac{\alpha_0}{3a^3}}.
\end{equation}
This equation gives the value for the magnetic susceptibility that is obtained when the Lorentz homogenization procedure is applied to the metamaterial with the ring magnetic polarizabilities $\alpha = \alpha_0(\omega_0^2/\omega^2-1)^{-1}$ already proposed in \cite{Marques-PRB-2002}, except for a small correction term, $2M_a/L\,+\,4M_c/L$, accounting for the effect of the closest rings. Actually, if such correction term is calculated by assuming a magnetic dipole approximation for the rings, it can be easily shown that it vanishes, thus giving exactly the Clausius-Mossotti formula for the susceptibility. Therefore, we can conclude that in the long wavelength limit, the transverse waves \reff{t-wave} correspond to the electromagnetic waves that are obtained from the local time-dispersive permittivity $\mu=\mu_0\{1+\chi(\omega)\}$. Conversely, in the short wavelength limit ($k_x\gg k_m$), they converge to the transverse magnetoinductive waves \reff{t-MI}. Furthermore, from $\vM_0 = A\vI_0/a^3$ and \reff{I0-equation} equation 
\begin{equation}
  \label{chi-equation}
  \vM_0 = \left\{ \frac{ja^3}{\mu_0\omega A^2}\ten{Z}(\vk, \omega)-\frac{1}{3}\right\}^{-1}\cdot \vH_0 = \ten{\chi}(\vk, \omega)\cdot \vH_0 \,
\end{equation}
can be obtained. Now in \reff{H-M-equation} $\vM_0$ can be replaced by \reff{chi-equation} leading to 
\begin{equation}
  \label{H-equation}
  \left\{\left(-k^2 + k_m^2\right) + \left(-\vk\vk + k_m^2\right)\cdot\ten{\chi}(\vk, \omega)\right\} \cdot \vH_0 = 0,
\end{equation}
which gives the same dispersion equation as \reff{d-r}. Therefore we can conclude that the non-local (i.e. time-and spatially-dispersive) magnetic permeability 
\begin{equation}
  \ten{\mu}(\vk, \omega) = \mu_0\left( 1 + \left\{ \frac{ja^3}{\mu_0\omega A^2}\ten{Z}(\vk, \omega)-\frac{1}{3}\right\}^{-1}\right)
  \label{mu-equation}
\end{equation}
provides a complete characterization of the metamaterial, accounting for all kind of waves propagating through it. In the long wavelength limit ($a|\vk|\ll 1$) all the exponential terms in $\ten{Z}(\vk,\omega)$ can be equated to unity and, from the aforementioned cancellation of the inductances between rongs of different orientation, all the off-diagonal terms of $\ten{Z}(\vk,\omega)$ vanish. Thus, $\ten{Z}(\vk,\omega)$ becomes an scalar and the magnetic susceptibility in \reff{chi-equation} reduces to the scalar $\chi(\omega)$ in \reff{t-EM}.

\begin{figure}
\centering
\includegraphics[width=0.8\columnwidth]{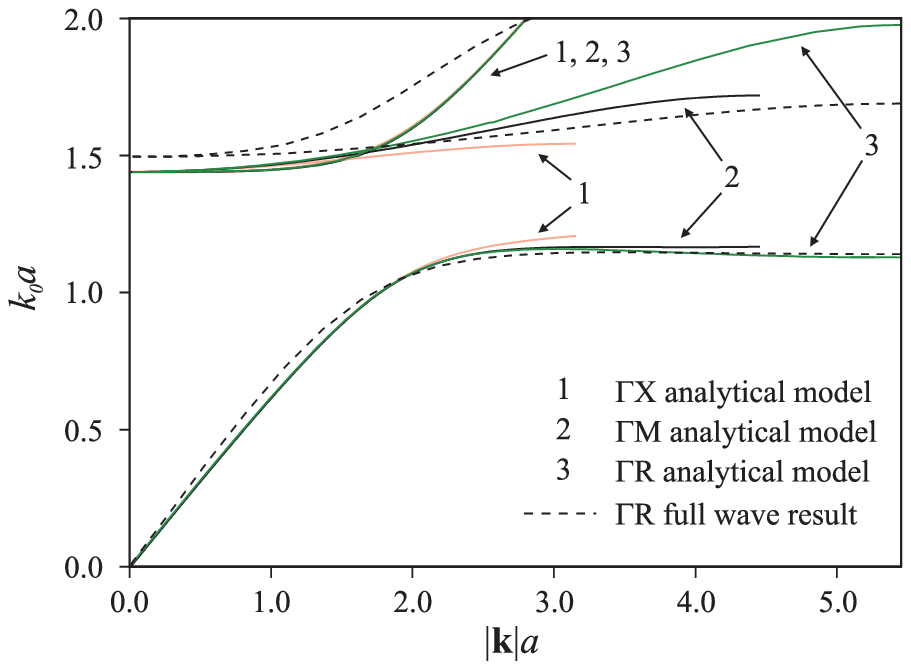}
\caption{\label{Fig2} (Color online) Dispersion diagram along $\Gamma$-X, $\Gamma$-M, $\Gamma$-R directions obtained from the analytical model (lines 1, 2, 3) and from a full-wave simulation (dashed line). The coordinates of selected points are $\Gamma= \left(0,0,0 \right)$, $X= \left(\pi / a,0,0 \right)$, $M= \left(\pi / a,\pi / a,0 \right)$, $R= \left(\pi / a,\pi / a,\pi / a \right)$.}
\end{figure}

\begin{figure}
\centering
\includegraphics[width=0.8\columnwidth]{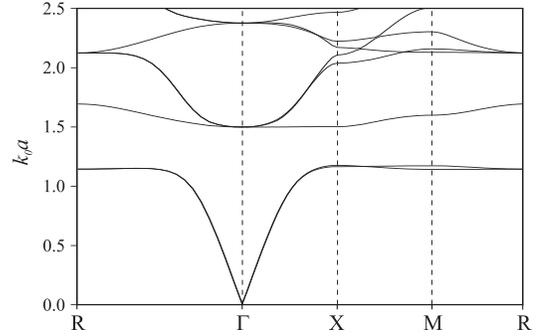}
\caption{\label{Fig3} Dispersion diagram along the closed path R-$\Gamma$-X-M-R obtained from full-wave simulation. Horizontal axis shows a projection of $\vk$ on the corresponding line on the boundary of Brillouin zone.}
\end{figure}

As a numerical example, we have studied the propagation of electromagnetic waves in a metamaterial formed by the simple cubic lattice of split ring resonators whose unit cell is depicted in Fig. 1b. The capacitance $C$ and self inductance $L$ were calculated following the ideas of Ref.\cite{Marques_AP_2003} but for the case of an SRR made of wires instead of planar strips \footnote{In this case the total capacitance of the SRR can be estimated as $C = \frac{1}{8} \pi R C_{pul}  = \pi^2 \varepsilon_0 R / \left( 4 \cosh ^{ - 1} \left( {\frac{{d^2 }}{{2r^2 }} - 1}\right)\right)$, where $C_{pul}$ is the per unit length capacitance of two parallel wires given in \cite{Jackson01}(p. 88). The self-inductance of the SRR was approximated by the self-inductance of a torus with large radius $R$ and small radius $r$ \cite{Jackson01}(p. 233): $L = \mu _0 R\left[ {\ln \left( {8\frac{R}{r}} \right) - 2} \right]$.}. The mutual inductances $M_{ij}^{\vn \vm}$ were calculated numerically using Neumann's formula including time retardation. The macroscopic permittivity was approximated by substituting the SRRs by planar conducting disks of the same external radius. Permittivity of such medium was calculated using static Lorentz homogenization theory \cite{Collin} and its value is $\varepsilon_r = 2.5$. Using Eq. \reff{d-r} and the first neighbors approximation, the dispersion characteristic of the electromagnetic modes supported by the metamaterial along different directions of the first Brillouin zone were calculated. The result for the geometry associated with $R=0.44a$, $r=0.005a$ and $d=0.03a$ (see Fig. 1b) is depicted in Fig. 2. It can be seen that the band structure is formed by three branches and contains a bandgap for all the depicted directions of $\vk$. For or $\vk$ along $\Gamma$-X the first and third branches corresponds to the transversal mode described by Eq. \reff{t-wave} and the second branch corresponds to the longitudinal mode described by Eq. \reff{l-wave}. Figure 2 also shows the high isotropy of the transversal mode even for moderate values of $\vk$, a fact that is expected from the tetrahedral symmetry of the system \cite{Baena_APL_2006}.

To assess the accuracy of the proposed analytical model, we have also numerically computed the \emph{exact} band structure of the aforementioned periodic material using the hybrid-plane-wave-integral-equation formalism introduced in \cite{HybridMTT}. The result of the numerical simulation is presented in Fig. 2 for the specific direction $\Gamma$-R, and in Fig. 3 for the closed path R-$\Gamma$-X-M-R. Good qualitative agreement between theory and simulation can be seen from Fig. 2. We think that the quantitative disagreement for high values of $k$ in the second branch can be attributed to the specific local field approximation considered in Eq. 2, which is strictly valid only for small values of $k$. In the case of the third branch, the disagreement is due to the proximity of the second resonance of the SRRs, which is not taken into account in the model. This effect is more visible in Fig. 3, where higher frequency branches are included. This figure also shows a complete electromagnetic band gap in the range $1.18 <k_0 a < 1.50$, in agreement with the hypothesis of a negative permeability in such frequency band. It may be worth noting that the effect of the substitution $k^{2}_{m} \rightarrow -k^{2}_{m}$ (or equivalently $\varepsilon_r \rightarrow - \varepsilon_r$) into \reff{t-wave-2} is the onset of a backward wave passband in the frequency range of the stop-band of Fig. 3, as well as the conversion of the passbands of Fig.2 into stop-bands. Therefore, it can be guessed that the proposed model will be also useful for the analysis of isotropic left-handed media made of SRRs and wires or any other elements providing a macroscopic negative permittivity (provided the conditions for the validity of the superposition hypothesis previously discussed in \cite{Marques/Smith-2004} are fulfilled). Work in this direction is in progress.

\begin{acknowledgments}
This work has been supported by the Spanish Ministerio de Educaci\'on y Ciencia under project TEC2007-68013-C02-01/TCM and by Spanish Junta de Andaluc\'ia under project P06-TIC-01368. The authors also thank Dr. Pavel A. Belov for useful discussions during the preparation of the paper.
\end{acknowledgments}

\end{document}